\documentclass[final]{aipproc}

\layoutstyle{6x9}

\SetInternalRegister\hbadness{8000} 

\begin{document}

\title 
      []
      {Modelling the ultraviolet/submillimeter spectral energy distributions 
of normal galaxies}


\author{Cristina C. Popescu \& Richard J. Tuffs}{
  address={Max Planck Institut f\"ur Kernphysik, Astrophysics Dept., 
  69117 Heidelberg\\
  email:Cristina.Popescu@mpi-hd.mpg.de; Richard.Tuffs@mpi-hd.mpg.de}
}

\copyrightyear  {2005}

\begin{abstract}
We give an overview of the factors shaping the ultraviolet (UV)/optical - 
far-infrared (FIR)/submillimeter (submm) spectral energy distributions (SEDs) 
of normal (non-starburst) galaxies. Particular emphasis is placed on the influence of the geometry of dust and stars on the propagation of light through the interstellar medium. Although strong constraints can be placed on the amount and large scale distribution of dust in disks from the appearance of the galaxies in the optical/UV range, this dust does not account for the observed amplitude and colour of the FIR/submm radiation. Additional, optically thick components of dust associated with the young stellar population on large and small scales are required to account for the complete UV/optical - FIR/submm SEDs. Self-consistent models for the calculation of SEDs of spiral galaxies are reviewed, and their predictions for the dust emission and the attenuation of starlight are compared and contrasted.
\end{abstract}
\date{\today}

\maketitle

\section{1. Introduction}
Historically, almost all our information about the 
current and past star-formation properties of normal galaxies has been based 
upon spatially integrated measurements in the
ultraviolet (UV), visible and near-infrared (NIR) spectral regimes. However, 
star-forming galaxies 
contain dust which absorbs some fraction of the emitted starlight, 
re-radiating it predominantly in the far-infrared (FIR)/sub-millimeter (submm) 
range.

In view of this, the measurement of current and past 
star-formation in galaxies - and indeed of the universe as a whole -
requires a quantitative understanding of the 
role different stellar populations play in powering the FIR/submm
emission. For this both optical and FIR/submm data need to be used, as they
contain complementary  information about the distribution of stars and dust.
The problem is very complex. Images of galaxies taken in the stellar light 
and dust emission show that galaxies are very inhomogeneous
systems, presenting both small scale and diffuse components, both in
the UV and in the FIR. For instance small scale structures 
show that part of the stellar light can be locally absorbed and re-radiated 
in the FIR. The diffuse emission shows that another part of the stellar 
photons can travel some distance in the disk before being scattered or 
absorbed. In order to quantitatively model these processes one needs to 
make a self-consistent calculation of the transport of radiation and 
its re-emission in galaxy disks. The challenge is to identify a
model which is sufficiently simple to make the problem tractable, but which 
can still predict the essential elements of observed SEDs and
morphologies. Here we review progress made towards identifying such models
for normal galaxies. Models for predicting SEDs for starburst galaxies are 
considered in the review of Dopita \cite{dop05}.

\section{2. What shapes the SEDs of galaxies?}

There are two fundamental factors to be considered here:
\begin{itemize}
\item the intrinsic emission spectrum produced by all stellar populations in 
a galaxy 
\item the propagation of photons through the ISM.
\end{itemize}

The intrinsic spectrum depends on the star formation history and
the metallicity evolution of a galaxy. We will not expand on this  here since
this is covered by the review talk of Leitherer \cite{lei05}.
The propagation of photons through the ISM depend on:
\begin{itemize}
\item the amount of dust and its distribution with respect to the stars
\item the optical properties of grains
\item the amount of neutral gas and its distribution with respect to the young stars
\end{itemize}

The neutral gas affects the line emission component of SEDs; this topic is
covered in detail in the talk of Kewley et al. \cite{kew05}. The optical
properties of grains, which mean their absorption and scattering properties
throughout the UV-submm range, depend on the composition and size 
distribution of the grains. This topic is covered in detail by the review 
talk of Dwek \cite{dwe05} (see also the talks by Li \cite{li05} and by 
Gordon \cite{gor05}), and is only briefly touched upon here, in 
Sect.~4. The amount of dust and its distribution with respect to the stars 
has received little attention until recently, even though the distribution of
dust is the single most important factor affecting the propagation of photons 
in galaxies. To illustrate this statement one only has to recall that a fixed 
amount of dust 
distributed on a scale of a few parsec around stars will have the same effect 
as 1000 times more dust distributed on kpc scales in the disk. Therefore this
review will mainly focus on the effect of the relative geometries of stars and
dust on the SEDs. Previous reviews related to this topic include: Calzetti
\cite{cal01a}, \cite{cal01b}, Popescu \& Tuffs \cite{poptuf02}, Kylafis \&
Misiriotis \cite{kylmis05}. 

We will start with very simple
geometries and increase the complexity until we have identify a minimum degree
of complexity that can account for observed broad-band 
UV/optical/FIR/submm SEDs. We will also consider the effect of these models 
on the UV/optical/FIR/submm surface brightness distributions. In terms of the
dust emission, we will place most emphasis on the FIR emission, rather than, 
for example, on the MIR emission. This is firstly because most of the energy 
absorbed by grains is re-radiated in the FIR, and secondly, because the FIR
colours of a galaxy depend on the strength of the radiation fields in a
galaxy, and therefore more directly constrain the propagation of photons in the
disk.  

Our entry point is to consider geometries with cylindrical symmetry, which are
essential for the description of disk galaxies. Spherical symmetry is a more
reasonable approximation for the description of dwarf galaxies 
(e.g. Galliano et al. \cite{gal03}) and starburst galaxies (e.g. Witt et
al. \cite{wit92}, Gordon et al. \cite{gor01}). 
The simplest model is the infinite slab/sandwich. A sandwich model, not
incorporating scattering, was used by
Disney et al. \cite{dis89} to investigate the attenuation-inclination 
behaviour of spiral galaxies. This work first emphasised the strong effect of 
the relative scaleheights of stars and dust on the attenuation. Another 
version of the sandwich model, this time including scattering, was used
to calculate the energy balance between the emission and re-emission of light 
in the pioneering work of Xu \& Buat \cite{xubua95}.  In the Xu \& Buat 
formulation there is only one free parameter, the face-on optical depth, 
which is adequate to account for the energy balance. Apart from its 
application to the integrated emission from galaxies (Buat \& Xu
\cite{buaxu96}, Xu et al. \cite{xu97}), this particular model 
was used by Xu \& Helou \cite{xuhel96} in the modelling of the large-scale dust
heating and cooling in the diffuse medium of M~31.
A common drawback, though, of models involving infinite slab/sandwich 
geometries is that they cannot predict the shape of the observed FIR/submm 
SEDs. Understanding the FIR colours is not only a matter of academic concern,
but also provides a further dimension to the predictive power of models, 
since the FIR colours directly probe the strength of the radiation fields, 
and, as we shall see, strongly depend on physical quantities of interest, 
such as SFRs.  

In order to also fit the FIR colours, one needs to consider more realistic
geometries, where by realistic we mean incorporation of 
finite disks, bulges and small scale structures:
\begin{itemize}
\item finite disks: stars and/or dust
\item bulges: stars only
\item small scale structures: stars and/or dust
\end{itemize}

Finite disks are usually described by double exponentials in both radial and
vertical direction. 
Bulges can be described by a variety of forms: de Vaucouleurs,
truncated Hubble, spherical with King profile/exponential. 
We note here that the exact choice
of bulge geometry has little effect on the shape of the globally integrated 
SED, provided that the bulk of the luminosity of the bulge is emitted
within an area much smaller than the disk. 
Small scale structures have been described as small scale dust
clouds/clumps (depending on terminology)
which, according to the model, may or may not be physically associated with 
young stars. As we shall see, this different treatment has strong consequences
on the prediction of the FIR colours. 
All or combinations of these geometrical components have been employed by 
the models introduced in the following papers: Kylafis \& Bahcall 
\cite{kylbah87}, Bianchi et al. \cite{bia96}, Bianchi et al. \cite{bia00a}, 
Xilouris et al. \cite{xil97}, Silva et al. \cite{sil98}, 
Granato et al. \cite{gra00}, Kuchinski et al. \cite{kuc98}, 
Popescu et al. \cite{pop00}, Matthews \& Wood \cite{matwoo01}.
All these models were used to account for the optical SEDs, but only 
three of them (Bianchi et al. \cite{bia00a}, Silva et al. \cite{sil98}, 
Popescu et al. \cite{pop00}) were used as the basis for a self-consistent 
calculation of both the stellar and dust emission SEDs. Recently, 
self-consistent calculations of attenuation and re-emission by dust grains in 
galaxies have also been done in the work reported by Baes et al. \cite{bae05} 
in this volume, and have started to be incorporated in population synthesis
models, such as those of Piovan et al. \cite{pio05} and Rocca-Volmerange
\cite{roc05}, also as reported in this volume.

\section{3. Models which account for the optical appearance of galaxies}

\begin{figure}[htb] 
\includegraphics[scale=1.]{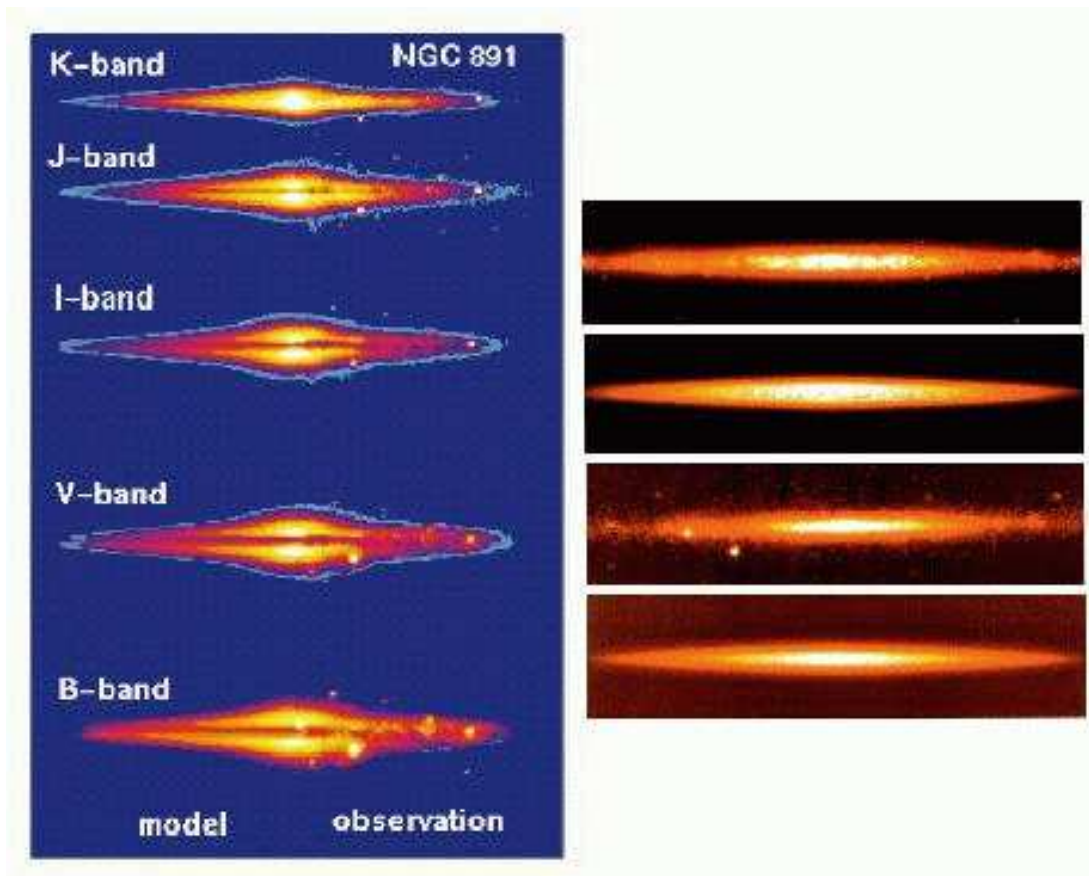}
\caption{Left panel: Images of NGC~891 in K, J, I, V, B bands (top to bottom)
  taken from Xilouris et al. \cite{xil98}. The left hand half of each galaxy 
  image is the model image and the right half is the real galaxy image
  (folded). Right panel: Comparison of R- and H-band Monte Carlo model images 
  of UGC~7321 (second and bottom rows) with real data at these wave bands (top
  and third row), taken from Matthews \& Wood \cite{matwoo01}.}
\end{figure} 

The challenge of models which account for optical SEDs is to
identify an intrinsic distribution of stars and dust which is consistent
with the observed integrated SEDs and surface brightness 
distributions.
To assess whether a particular choice for the geometry 
of stars and dust is consistent with observations, calculations
of the transfer of radiation through the galaxy must be done. Thus, models
must incorporate a radiative transfer code. Different
radiation transfer codes have been discussed in the review talk of Kylafis
\& Xilouris \cite{kylxil05}. 

The model of Xilouris et al. \cite{xil97} derived the
relative distribution of stars and dust by fitting optical/NIR images of
galaxies, assuming the simplest, yet realistic, distributions of stars and 
dust that could be used to describe such systems (i.e., exponential disks plus
a de Vaucouleurs bulge; see Fig.~1, left panel). This 
work was done for edge-on galaxies, which allows
not only the scalelength but also the scaleheights of dust and stars to be 
extracted. We should also mention that, since opacities were independently 
derived at each wavelength, the extinction law was extracted for each modelled 
galaxy, giving some constraints on the dust model. Apart from the detailed 
knowledge of the distributions of stars and dust in individual galaxies, some 
general trends became evident (Xilouris et al. \cite{xil99}), namely:
all the galaxies modeled were able to reproduce Milky Way extinction laws; 
the optically emitting stars have a scaleheight which is about twice that of
the dust; the dust scalelength is about 1.4 times larger than that of the 
stars and the dust is more radially extended than the stars. 
This inference about the relative radial extent of stars and dust
can in principle be tested through FIR observations of the dust emission
 from the regions beyond the stellar disk. This has been done in the case
 on NGC~891 (Popescu \& Tuffs \cite{poptuf03}; see also Fig.~2 from the 
review of Tuffs \& Popescu \cite{tufpop05}). 

Synthetic images have also been compared with observed images by 
Matthews \& Wood
\cite{matwoo01} (see Fig.~1, right panel), but in this work the geometry was fixed and 
only the face-on optical depth was varied. 


\section{4. Models which account for the entire optical/FIR/submm SEDs}

The challenge of models which account for the entire optical/FIR/submm SEDs 
is to identify an intrinsic distribution of stars and dust which is consistent
with the observed integrated SEDs and surface brightness 
distributions, both in the optical and FIR/submm ranges. In addition to
the radiative transfer code, such models should also incorporate a 
technique to calculate the dust emission, as outlined below:

For a given spatial distribution of stars and dust, we can calculate
the energy densities $u_{\lambda}$ at each position in the galaxy through 
a radiative transfer calculation, which needs to take into account both 
absorption and scattering. Then, for a given grain model (which in this
context means a given grain absorption/emission efficiency $Q_{abs}$) and a 
given distribution $N(a)$ in (spherical) grain size $a$, the energy balance
between emission and absorption is given in terms of a
probability distribution $P(a,T)$ in the temperature $T$ of each grain by
the following equation:

\begin{eqnarray}
\int_0^{\infty} Q_{abs}({\lambda},a)\,u_{\lambda}\,
d{\lambda} = \frac{4{\pi}}{c}
\int_0^{\infty} Q_{abs}({\lambda},a)\,d{\lambda}
\int_0^{\infty}B_{\lambda}(T)\,P(a,T)\,{\rm d}T \nonumber
\end{eqnarray}\\
Having solved for $P(a,T)$ (see for example Guhathakurta \& Draine
\cite{guhdra89}), the dust emission $F_{\lambda}$ can be
calculated at each position in a galaxy at distance $d$ using:
\begin{eqnarray}
{F_{\lambda}} & = & \frac{1} {d^2}\int_{a_{\rm min}}^{a_{\rm max}} 
N(a)\,{\rm d}a\,{\pi}\,a^2\,Q_{\rm abs}{(\lambda},a)
\int_0^{\infty}B_{\lambda}(T)\,P(a,T)\,{\rm d}T \nonumber  
\end{eqnarray}\\
where $a_{min}$ and $a_{max}$ are the minimum and maximum grain sizes present
in the distribution.  

\begin{figure}[htb] 
\includegraphics[scale=1.0]{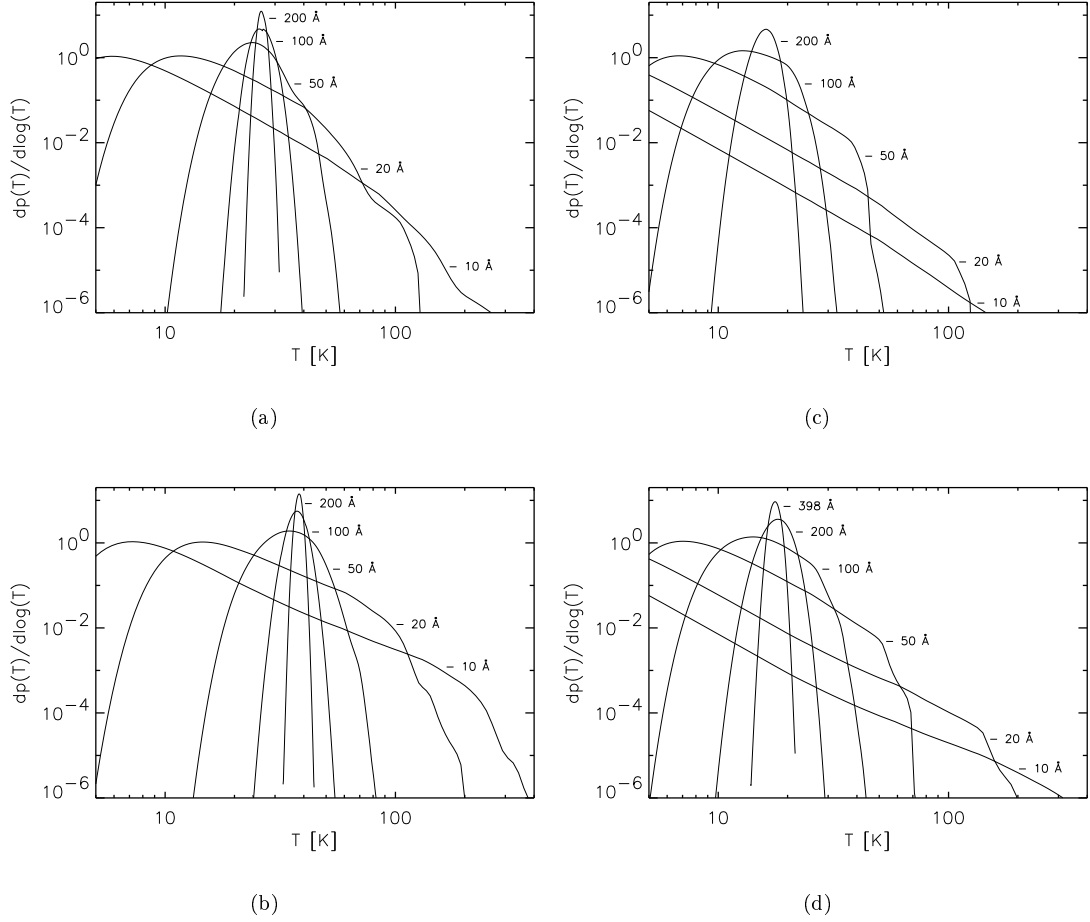} 
\caption{Examples of temperature distribution for spherical dust grains 
(with radii 10, 20, 50, 100, 200, \& 398\,\AA) 
embedded in the diffuse radiation 
field of NGC~891, taken from Popescu et al. \cite{pop00}: 
a) silicate grains in the centre of the galaxy
($R=0$\,kpc, $z=0$\,kpc); b) graphite grains in the centre of the galaxy;
c) silicate grains at the edge of the galactic disk 
($R=15$\,kpc, $z=0$\,kpc); d) graphite grains at the edge of the galactic 
disk.}
\end{figure}

A probability distribution of temperatures arises because grains are 
impulsively heated by photons. Only when the typical interval between
photon hits is much shorter than the timescale for the energy deposited
in each photon hit to be radiated does the probability distribution
tend towards a delta function - the case of grains emitting at their
equilibrium temperature. In general grains smaller than a certain critical size
(depending on the strength and colour of the radiation field and composition
of the grain) emit stochastically, and grains bigger than this critical size 
emit at equilibrium temperature. As the radiation fields are increased, 
grains of progressively smaller sizes radiate at equilibrium temperature. 
Conversely, as radiation fields are decreased, grains of progressively larger 
sizes will radiate stochastically. These trends are illustrated in Fig.~2.

\subsection{4.1 Choice of intrinsic distributions of stars and dust}

\subsubsection{The old stellar disk and associated dust}

\begin{figure}[htb] 
\includegraphics[scale=0.85]{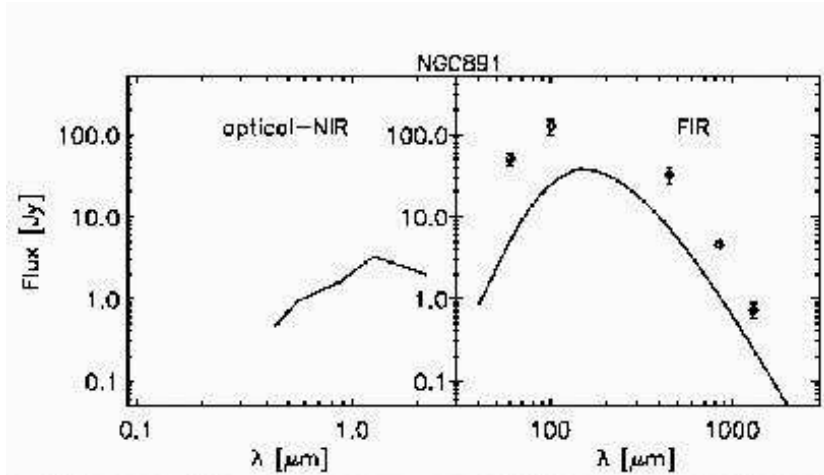}
\caption{The FIR/submm SED of NGC~891 predicted for the old stellar population
  and associated dust (taken from Misiriotis et al. \cite{mis01}). It is clear 
that the prediction (solid line) for the FIR/submm SED falls well short of 
the observed measurements (symbols).}
\end{figure}

The obvious starting point in identifying a spatial distribution of stars and 
dust that
can account for the entire optical/FIR/submm SED is to test whether the
distributions which are consistent with the observed optical SEDs can
also account for the FIR/submm SEDs. 
We will use the model of Xilouris et al. \cite{xil97} to address this 
question as it is the
only model that fits the geometry of stars and dust as seen in the optical, 
and so should provide 
the best representation of
the intrinsic distributions of the stars and dust in spiral
galaxies that can be probed by optical extinction measurements.

To provide a benchmark, we will in the first instance
ignore the heating of dust by UV photons, since these
could not be included in the analysis of the edge-on systems considered
by Xilouris et al. For the FIR emission calculation we use
the standard dust model of graphite/silicate from Draine \& Lee \cite{dralee84}
(and with optical constants specified by Laor \& Draine \cite{laodra93}). 
This is consistent with the Milky Way type extinction 
law derived for these galaxies. The results are shown in Fig.~3. It is clear 
that the prediction for the FIR/submm SED falls well short of the observed 
measurements. In terms of luminosity the shortfall is by a factor of 5. This 
shows that, according to these models, the old stellar population alone cannot 
account for the dust emission.

\subsubsection{Adding the young stellar disk and associated dust}

Lets now add the UV emitting stellar population, but keep the amount and 
distribution of dust the same. This was considered as a new geometrical 
component in the model of Popescu et al. \cite{pop00}. The UV emissivity was 
assumed to be distributed in a thin
disk with the same scalelength as the B band disk, but with an assumed scale
height of 90 pc, close to that of the Milky Way (\cite{mihbin81}). This is 
smaller than the scale heights for the optically
emitting stars and for the dust distribution derived in the model of 
Xilouris. The spectral distribution of the
UV emission was fixed according to the predictions of a population synthesis
model, and its amplitude parameterised in terms of a SFR. Now it is easy to
raise the level of the predicted FIR emission to the observed level 
by simply adjusting the SFR. (see Fig.~4).

\begin{figure}[htb] 
\includegraphics[scale=0.53]{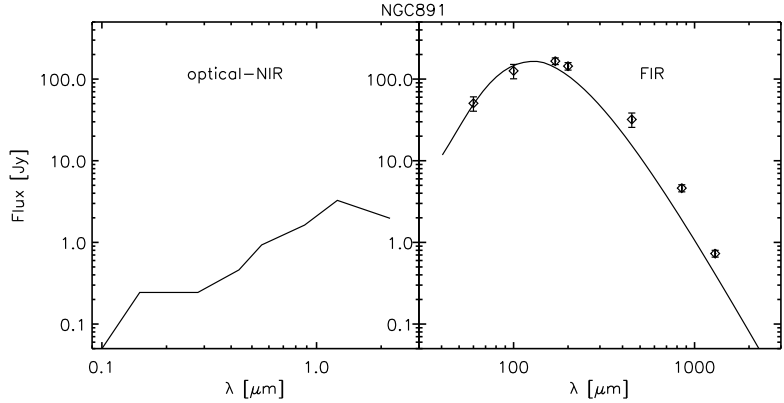}
\caption{The FIR/submm SED of NGC~891 predicted for the old stellar population
  and associated dust, supplemented by the young stellar disk (taken from
  Popescu et al. \cite{pop00}). It is obvious that the model
  prediction (solid line) for the submm emission falls short of the observed 
  measurements (symbols).}
\end{figure}

It is obvious from Fig.~4 that the FIR-submm colours are not fitted,
in the sense that the model prediction for the submm emission falls short
of the observed emission. This suggests either that a further component of
dust is present which is not constrained by the optical data, or that
the ratio of the submm emissivity to the FIR emissivity is higher than that 
predicted by the standard Draine \& Lee \cite{dralee84} dust model. 
The latter possibility
was recently investigated by Alton et al. \cite{alt04} and Xilouris
\cite{xil05}, who found a submm emissivity enhanced by a factor of
4. This result was obtained by comparing the 
visual optical depth derived from Xilouris's model with the submm 
emission, assuming that the same grains emitting in the submm are responsible 
for the optical extinction. However, this assumption is only true if 
self-shielding effects of grains are not present. In fact, real galaxies do 
contain 
optically thick components, which makes it inevitable that the submm
emission will always be higher than predicted on the basis of optical
extinction alone. Therefore the solutions obtained by Alton et al. \cite{alt04}
and Xilouris \cite{xil05} for the submm emissivity can only be treated as upper
limits. We should also note that there is no direct observational evidence of
enhanced submm emissivity in the {\it diffuse} ISM. It has only been possible 
to investigate submm emissivities of grains in discrete clouds, where 
the brightness of the dust emission from the clouds can be compared with the 
extinction of background stars. Although enhanced
submm emissivities have been found in these investigations, this is not
unexpected since the clouds shield the grains from the 
diffuse UV radiation field, allowing ice mantles to form. Thus,
such grains are probably not typical of grains in the bulk of the
volume of the galaxy, and, as already noted, are in any case 
not probed by global extinction studies due to the low filling factor 
of the clouds. We also note that there are other constraints on 
the wavelength dependence of grain emissivity (see Li \cite{li05}) which
would tend to favour a steep decline of emissivity with increasing 
wavelength in the submm. A steep decline in emissivity would not however lead 
to the enhanced ratio of the submm to the FIR emissivity needed
to explain the FIR/submm colours of galaxies in the absence of a further
dust component. 
\begin{figure}[htb] 
\includegraphics[scale=0.95]{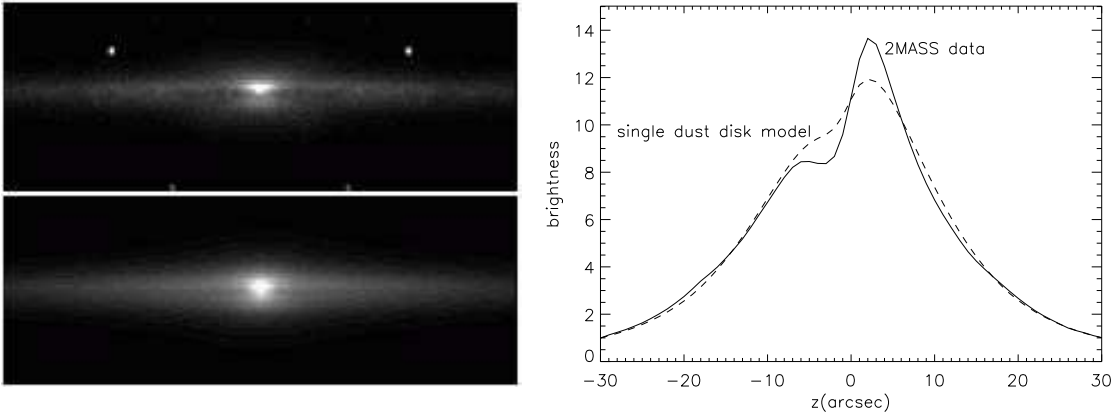}
\caption{Left upper panel: The 2MASS K-band image of NGC~891 (folded). Left
  lower panel:
  The simulated K-band image produced using the solution of 
Xilouris et al. \cite{xil98}, containing a single dust disk associated with the
old stellar population. Especially in the central region, the dust lane seen 
in the 2MASS image is too strong to be accounted for by the single dust 
disk in this model. This is quantified in the right hand panel, which shows the
  average brightness of the data and of the model (in arbitrary units) within 
50 arcsec of the centre, plotted versus vertical position.}
\end{figure}
In conclusion, there is no independent observational evidence
that an interstellar medium composed entirely of {\it diffuse} dust, 
as probed by 
the optical extinction studies, could have enhanced submm emissivity.

By contrast, there is strong independent observational evidence for a further
dust component to explain the observed FIR-submm colours of galaxies.
The geometrical model that we presented previously has no dust component 
associated with the young stellar population, yet we know that such a 
component exists. We know that dust is associated with the CO emitting layer 
and we know that this dust is associated with the spiral arms where most of 
the young stars are formed. The only question is how to prescribe the 
distribution of this dust. In order to mimic the effect of this component of 
dust associated with the young stellar population, and in the absence of 
direct observational constraints, Popescu et al. \cite{pop00} introduced a 
second dust disk, having the same scale height as the young stellar disk.  A 
second disk of dust associated with the molecular layer was also considered 
by Bianchi et al. \cite{bia00a} for the modelling of NGC~6946. Of course, in 
reality the dust
distribution is much more complex due to the spiral arm structure and
the (at least in part) clumpy nature of the dust associated with the young 
stellar population. Because of the presence of clumps, some fraction of the 
additional dust will be subject to strong self-shielding, and will hardly be 
seen at all even in the K band images. This may be one reason why the dust 
lanes observed on K-band images of edge-on galaxies are not much more 
prominent than the 
dust lanes predicted from the single dust disk model of Xilouris et al. 
(see the comparison of Dasyra et al. \cite{das05}).  Nevertheless, the 
dust lane seen on the K-band images
(Fig.~5) is, particularly in the central regions, too strong to be accounted 
for by the single dust disk model. This provides a hint that the additional 
dust layer is more concentrated towards the centre of the galaxy, following 
the distribution of molecular gas. 
\begin{figure}[htb] 
\includegraphics[scale=0.53]{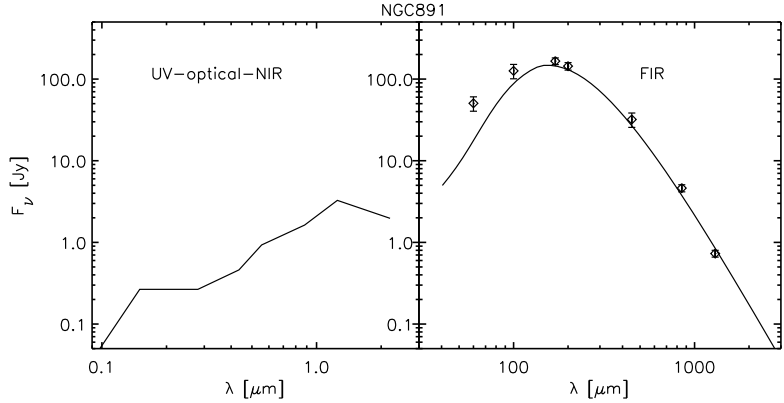} 
\caption{The FIR/submm SED of NGC~891 predicted for the old stellar population
  and associated dust, supplemented by the young stellar disk and associated
  dust (adapted from Popescu et al. \cite{pop00}). It is obvious that the 
  model prediction (solid line) for the shorter FIR wavelengths falls short 
  of the observed measurements (symbols), indicating the need for a warmer 
  component of emission. }
\end{figure}
Finally, we note that the prevalence of 
clumps in the dust associated with the young stellar population may reduce 
the dust mass required to account for the submm data, since, as we have 
already mentioned, dust in optically thick clumps may have an enhanced 
submm emissivity compared to dust in the diffuse inter-clump medium. Thus, the
dust mass derived by Popescu et al. \cite{pop00} for the additional dust
component should be treated only as an upper limit.
In this sense, elements of both the enhanced emissivity hypothesis and the 
additional dust hypothesis may be needed for a realistic description of the 
SEDs of normal galaxies.

Despite the complex structure of the dust associated with the young
stellar population, the simplification of placing it in a common thin
disk with the young stars is a reasonable approximation to predict the
optical appearance of galaxies (Misiriotis et al. \cite{mis02}), as well as 
the overall energy balance and FIR and submm radial profiles of spiral
galaxies (as will be shown here). The effect on the predicted FIR/submm SEDs 
of including the second disk of dust in the radiative transfer
calculation is shown in Fig.~6. Although the longer wavelengths points are 
now well fitted, the predicted SED is not broad enough to fit the short 
wavelengths as well. We are clearly missing a warmer component of emission.

\subsubsection{Adding the clumpy component of dust with embedded young stars}

The lack of a warmer component of emission in the FIR SEDs predicted by the
models should not come as a surprise since we have not yet included the 
localised emission components associated with the star-forming regions. We 
know that galactic star formation regions have warm dust emission from the 
local absorption of non-ionising UV produced by the young massive stars.
In other words, in addition to the diffuse large scale dust disks, the model 
must also incorporate a clumpy component of dust, whose spatial distribution 
is correlated on scales of a few parsecs with the HII regions. 

\begin{figure}[htb] 
\includegraphics[scale=0.8]{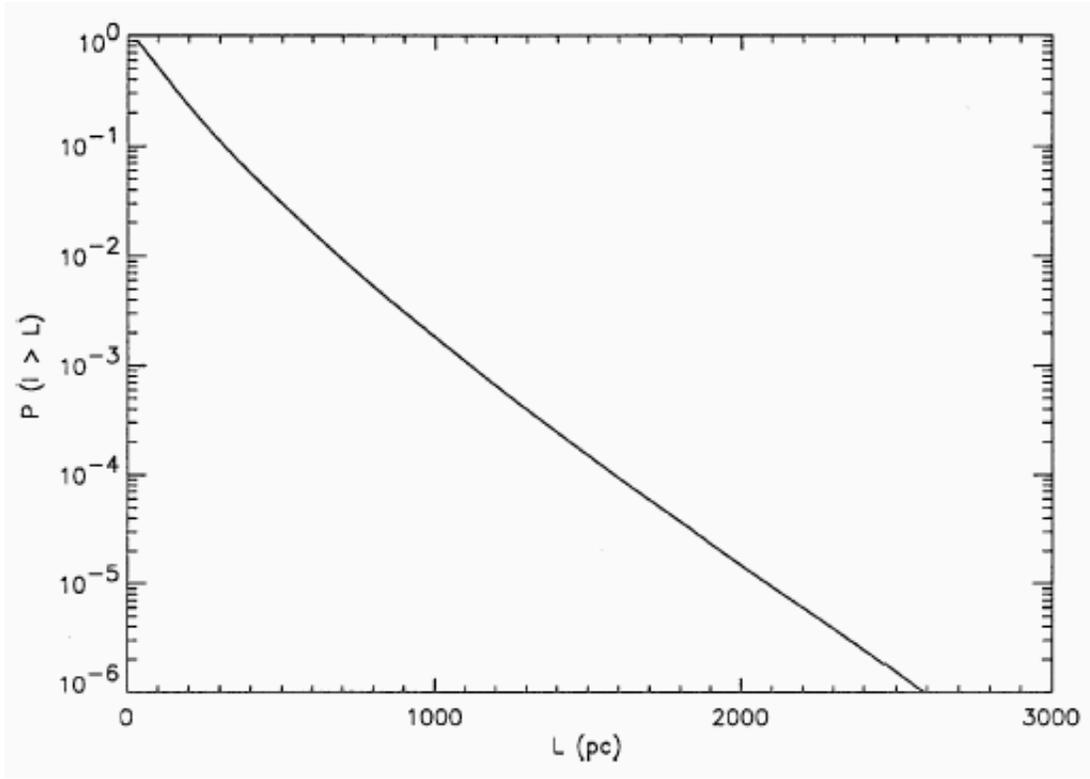} 
\caption{Distribution of the distance travelled by UV photons from 
 their parent OB association, before absorption, taken from Sauty et
  al. \cite{sau98}. The median distance is 120 pc, and the mean distance is 440
  pc.}
\end{figure} 
\begin{figure}[htb] 
\includegraphics[scale=0.68]{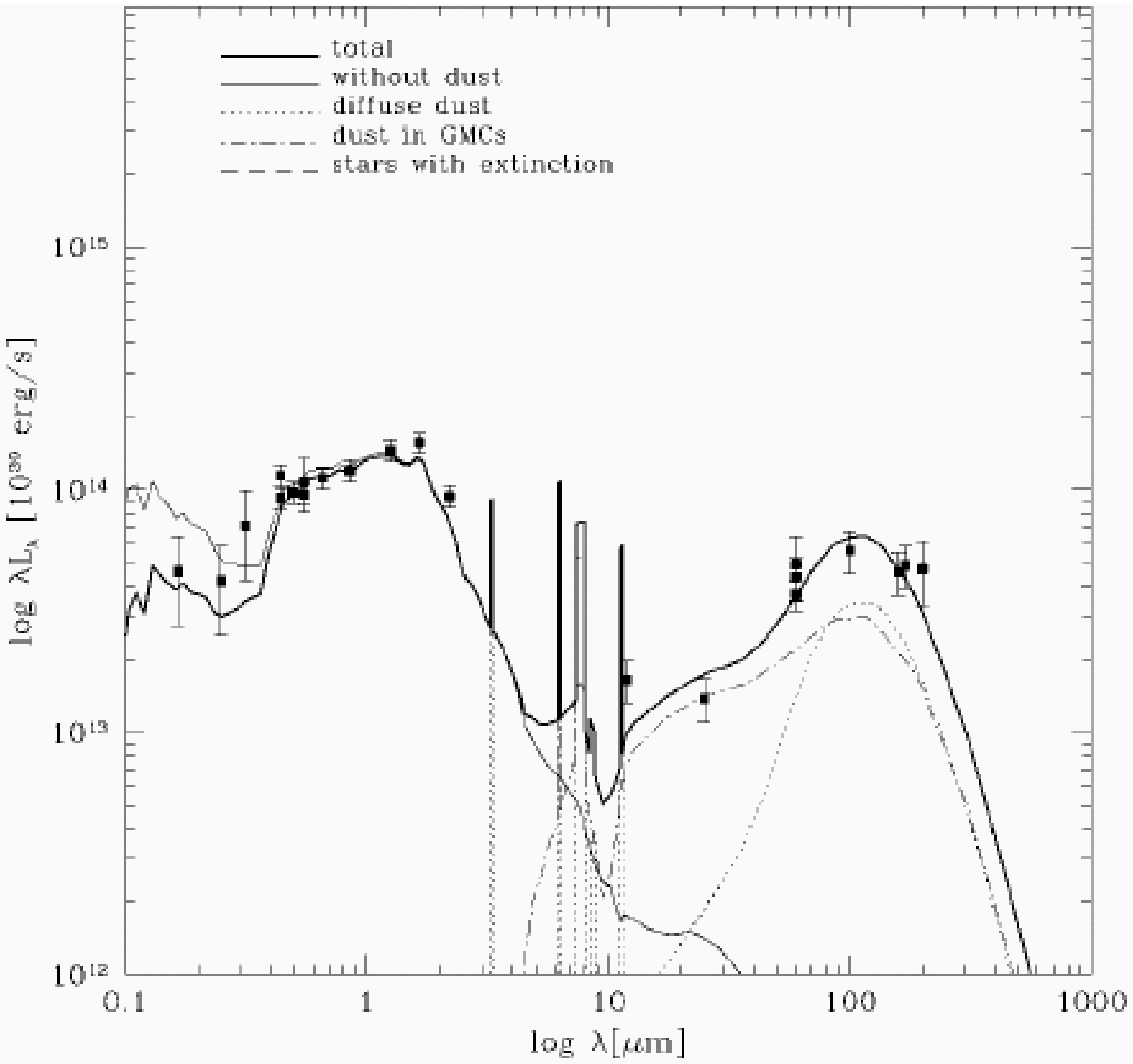} 
\caption{Fit to the SED of NGC~6946, taken from Silva et al. \cite{sil98}.}
\end{figure}

Ideally one would like to have radiation transfer calculations on scales
ranging from pc to kpc, to properly understand the propagation of photons on
both short and large scales. The closest that has been achieved is the
radiative transfer calculation of Sauty et al. \cite{sau98}. They 
calculated the transfer of UV radiation in a interstellar medium consisting 
of star-forming molecular clouds, visualised at a resolution of 12\,pc, as 
well as a diffuse atomic medium extending  up to 12\,kpc. They 
derived the distribution of the distance travelled by UV photons
before absorption. Fig.~7 shows that half of the radiation is 
absorbed within 
the first $\sim 10$ resolution elements out to a radius of $\sim120$\,pc. 
This calculation
cannot tell us, though, what happens within the first cell, which is larger 
than the size of most HII regions and their immediate surroundings. From an 
infrared point of view this means 
that the calculation is not sensitive to the absorption on the shortest 
scales, and therefore is not sensitive to the warmest grains. It is thus 
preferable to consider the local absorption and re-emission of UV photons 
separately from these processes on galactic scales.  This approach was adopted
both by Silva et al. \cite{sil98} and by Popescu et al. \cite{pop00} (see also
Charlot \& Fall \cite{chafal00}).

\begin{figure}[hbt] 
\includegraphics[scale=0.53]{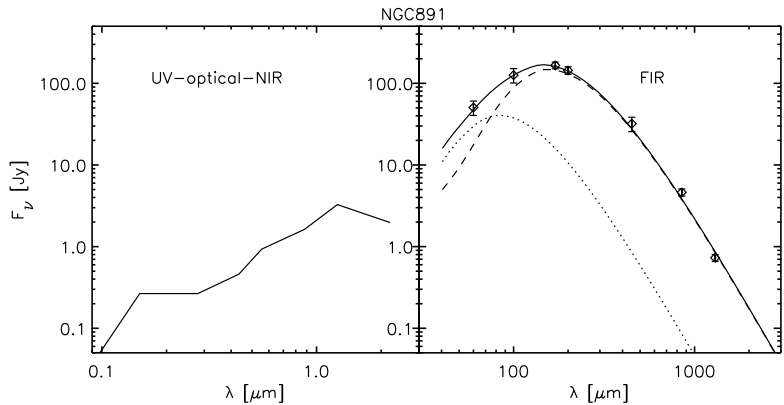} 
\caption{The final fit to the FIR/submm SED of NGC~891, taken from Popescu et
  al. \cite{pop00}. The total predicted SED is given with the solid line, the
  diffuse component with the dashed line and the clumpy component with the 
  dotted line. The observed data are given as symbols.}
\end{figure}

The basic idea is to consider that some fraction $F$ of the 
non-ionising UV is locally absorbed in star-forming complexes and only a 
fraction $(1-F)$ will go in the diffuse young stellar disk. 
If this clumpy component is added, the whole FIR/submm SED can indeed 
be fitted. This can be seen for both the model of Silva et al. \cite{sil98}
in Fig.~8 and for the model of Popescu et al. \cite{pop00}
in Fig.~9.  
Both figures show the contribution of the diffuse and localised components of 
the FIR emission. We note that in the model of Silva et al. \cite{sil98}
the diffuse and localised components peak at similar wavelengths, whereas in the
model of Popescu et al. \cite{pop00} they do not. This difference is due to 
the different treatment of the localised component. The model of Popescu et 
al. \cite{pop00} has a fixed empirically determined template for the FIR
emission from star-forming complexes, specified by just one parameter (its
amplitude - the factor $F$). The model of Silva et al. \cite{sil98} 
incorporates a separate radiative transfer calculation for an idealised 
star-forming complex, visualised as a molecular cloud with a central embedded 
source and parameterised by the radius and lifetime of the clouds. Thus,
whereas the model of  Silva et al. allows for the possible presence of cold 
dust emission from the optically thick parts of their  localised component, 
the model of Popescu et al.  considers only the warm dust emission from this 
component, and instead includes the cold dust in the second disk of dust 
which mimics the distribution of molecular material. This conceptual
difference will however not have a major effect on the
results obtained by the two techniques. The major difference is that one 
model (Popescu et al.) constrains the geometries of stars 
and dust from fitting the optical images, and has therefore only a few 
(in fact just three) free parameters needed to fit the FIR SED, while the other
model (Silva et al.) has more free parameters, partly because it also makes 
a calculation for the photometric evolution of a galaxy, but also because it 
has to define the geometries with further free parameters.

\subsection{4.2 Testing the models}

We have seen that a realistic model which can predict the observed
optical and FIR/submm SEDs needs the following components:
\begin{itemize}
\item stellar disk of old stars and associated dust disk
\item stellar disk of young stars and associated dust
\item bulge of old stars
\item clumpy component of dust with embedded young stars
\end{itemize}
We have also seen that such a model can also predict the optical appearance of
galaxies. But a more stringent test of these models is to see if they can also
predict the appearance of galaxies in the FIR/submm.  Such a test has been done
in just one case (for NGC~891; Popescu et al. \cite{pop04}) by comparing the 
images at 170 and
200\,${\mu}$m taken with the ISOPHOT instrument on board the Infrared Space
Observatory (ISO) with the model predictions of Popescu et al. \cite{pop00} 
(see Fig.~10). The comparison between the observed maps and the simulated maps 
for the diffuse component (at 170 and 200\,${\mu}m$) show a remarkable 
agreement. To search for small differences between the model and the 
observations, not 
detectable in the maps due to the high dynamical range of the displayed data, 
residual maps of the difference between the observed and the simulated maps 
are also shown. The residuals show the localised sources, which account for 
about 10$\%$ of the total emission at these wavelengths, in accordance with 
the prediction of the model. Another comparison is between observed and
predicted  radial profiles (see Fig.~11, left panel). Here again one can see 
a very good agreement. Again, the small excess 
emission in the observed profile is in agreement with the model prediction for
localised sources. To date, NGC~891 is the only galaxy for which an intrinsic
distribution of dust and stars could be found from a self-consistent modeling 
and which simultaneously accounts for both the optical and FIR morphologies. 

\begin{figure}[htb] 
\includegraphics[scale=0.75]{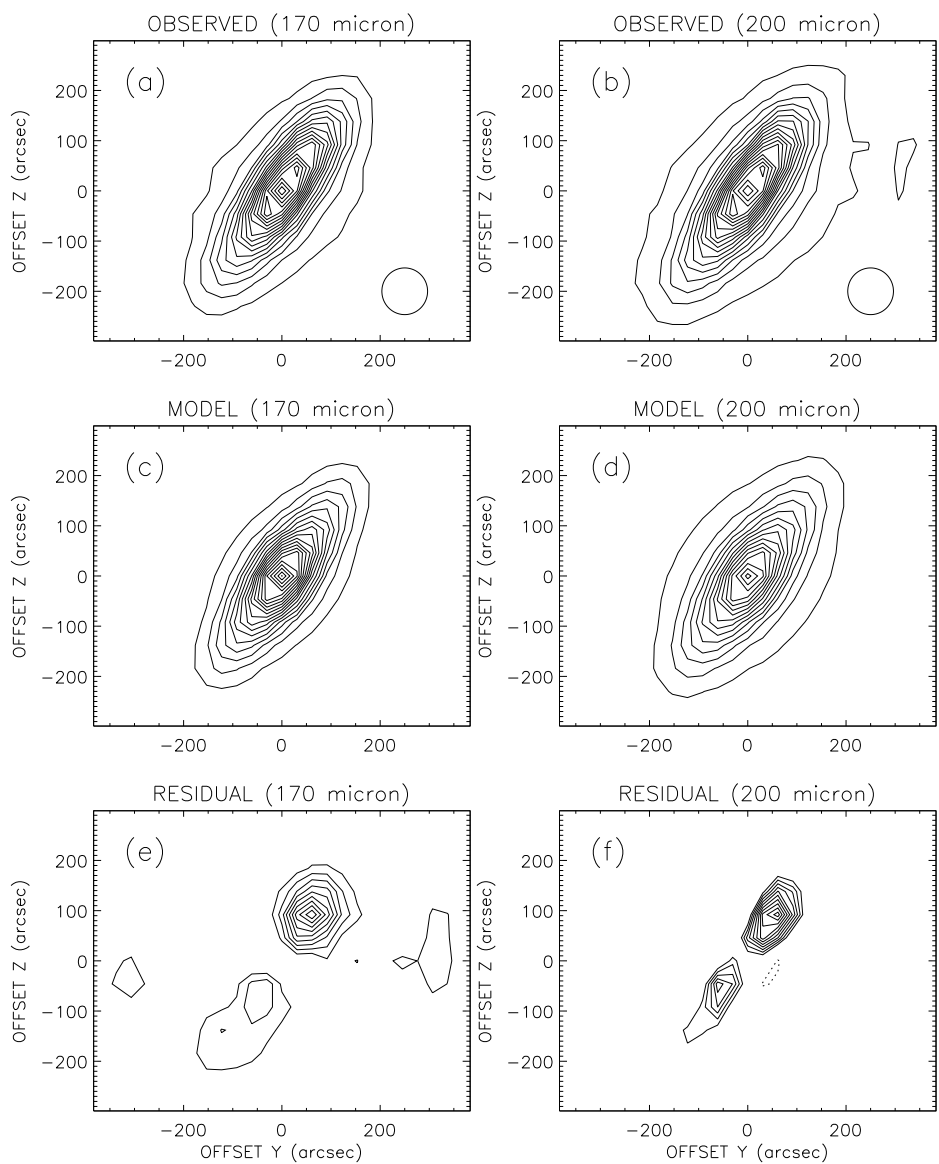} 
\caption{Comparison between the observed and predicted FIR maps of NGC~891,
  taken from Popescu et al. \cite{pop04}. The top rows show the observed maps
  taken with the ISOPHOT instrument on board ISO,
  the middle rows show the predicted maps for the diffuse component, and the
  bottom rows show the localised sources, obtained as residuals between the
  observed maps and predicted maps of the diffuse component. The circles from
  the top rows indicate the footprint (to the FWHM) of the ISOPHOT beam at 170
  and 200\,${\mu}$m. Y and Z are the spacecraft coordinates.}
\end{figure}

\begin{figure}[htb] 
\includegraphics[bb=65 276 375 645,scale=0.52]{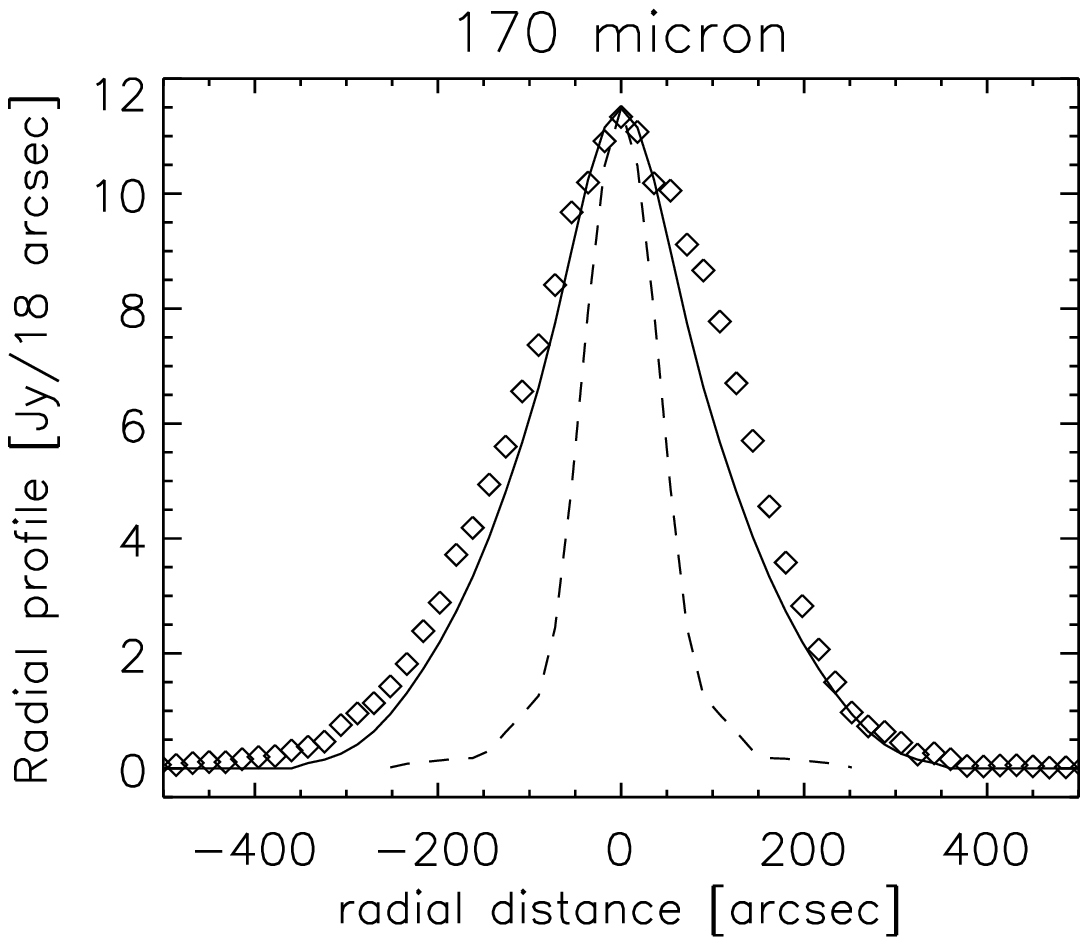}
\includegraphics[scale=0.80]{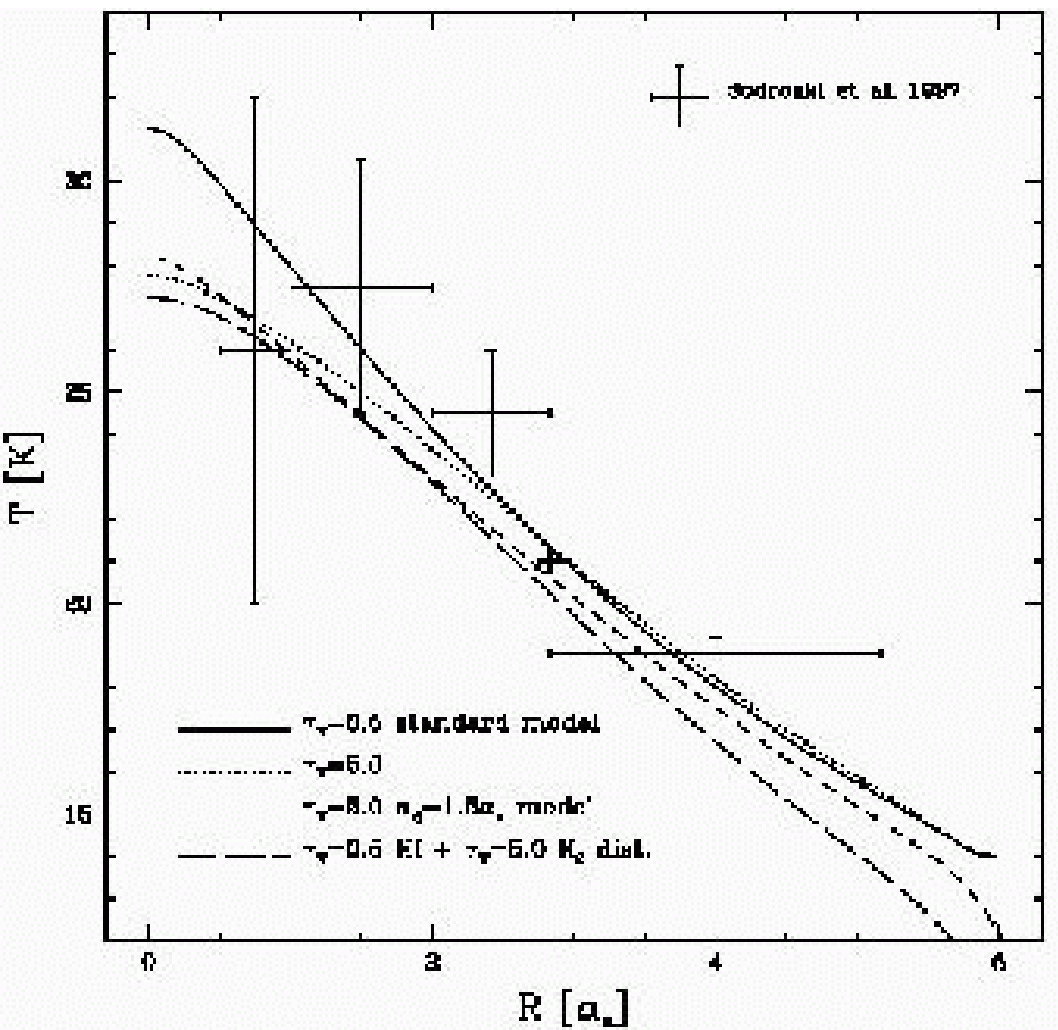}
\caption{Left: Comparison between the observed (symbols) and predicted 
(solid line) FIR radial profiles of NGC~891, taken from 
Popescu et al. \cite{pop04}. The predicted profiles are only for the diffuse
component. The small excess emission in the observed profiles is due to
localised sources. The
observations were done with the ISOPHOT instrument on board ISO. The
profiles were obtained by integrating the emission parallel to the minor axis
of the galaxy for each bin along the major axis. The sampling of the profiles
is 18 arcsec. The dotted lines represent beam profile. Right: Comparison 
between the predicted radial gradient of temperature in NGC~6946 and the 
observed one in the Milky Way, taken from Bianchi et al. \cite{bia00a}.}
\end{figure}

\begin{figure}[htb] 
\includegraphics[scale=0.77]{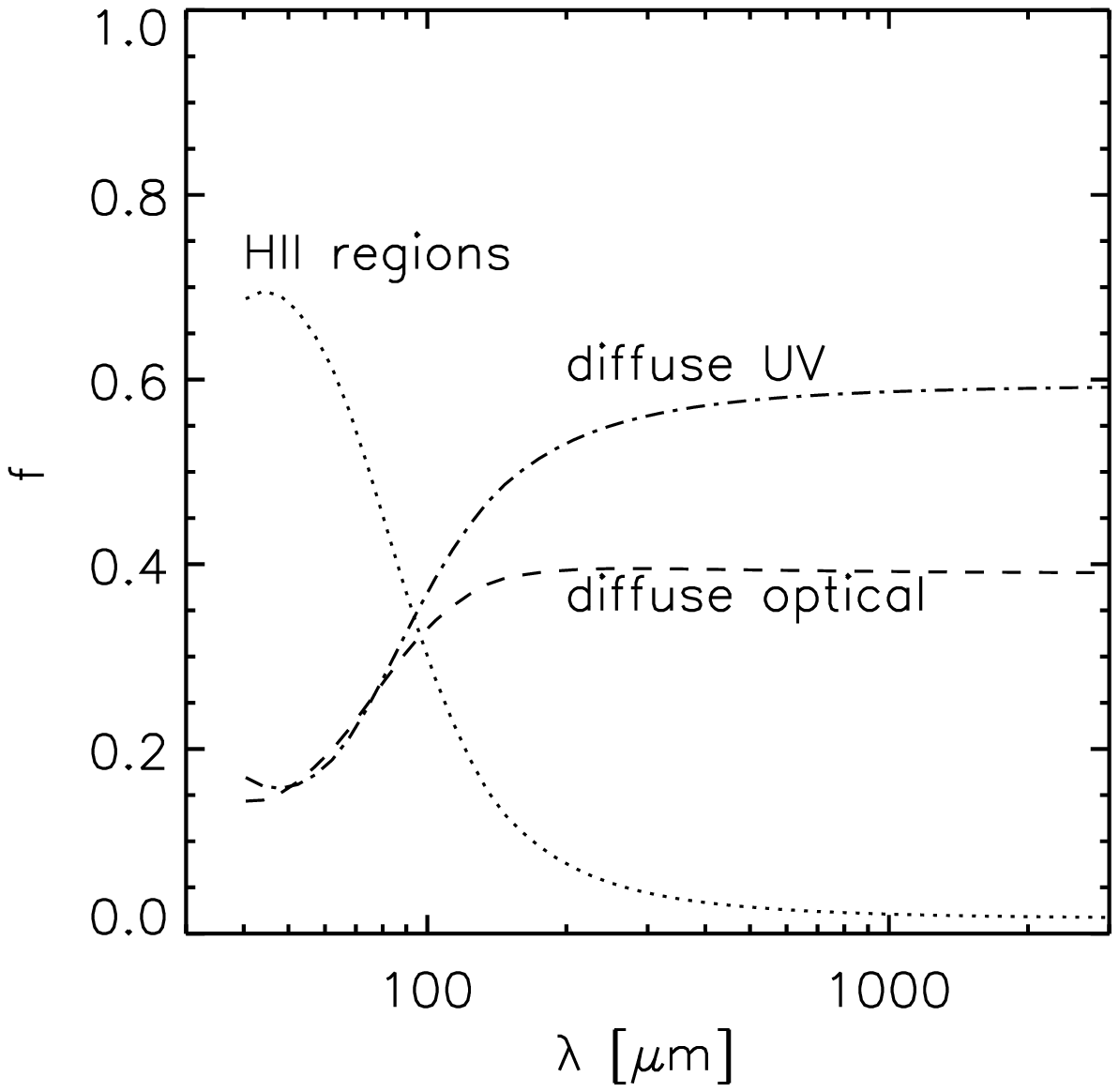} 
\caption{The fractional contribution of different stellar components to the 
dust emission, as a function of FIR/submm wavelength, taken from Popescu et
al. \cite{pop00}.}
\end{figure}

A related test utilising profiles rather than integrated quantities has been
done for the radial temperature distribution, namely by comparing that 
predicted by the model of Bianchi et al. \cite{bia00a} for NGC~6946 with that 
observed for the Milky Way (see Fig.~11, right panel).

It will be desirable to compare predicted and observed FIR/submm maps
for a larger number of galaxies. Work in this direction has also been 
started by Baes et al. \cite{bae05}, this conference.

\subsection{4.3 The origin of dust heating}

There has been a long standing question in the literature of what is the
contribution of different stellar populations to the heating of dust. This
question is equivalent to the question of the relative attenuations (averaged
over viewing angle) of individual stellar populations, for example in the 
bulge and in the old and young stellar disks. The self-consistent models of
optical/FIR/submm SEDs provide direct means to address this question.

In the modelling of the face-on spiral
NGC~6946, Bianchi et al. \cite{bia00a} found that it is the old stellar 
population that predominantly powers the dust emission, with only 40$\%$
of the emission powered by UV photons. However this 
result assumed that more than 50$\%$ of the absorbed 
UV radiation is channeled directly into the MIR, whereas self-consistent
calculations of this percentage show that most of the UV luminosity is
actually re-emitted in the FIR. For the same galaxy both 
Silva et al. \cite{sil98} and Sauty et al. \cite{sau98} found that most 
of the dust heating comes from the young stellar population. 
In the modelling of the edge-on galaxy NGC~891 Popescu et al. \cite{pop00} 
also found that it is the young stellar population that predominantly heats 
the dust, powering about 70$\%$ of the total dust luminosity. 
Furthermore, because the dust emission from the different components has 
different characteristic temperatures, one can determine the fractional 
contribution of the different stellar populations as a function of infrared 
wavelength. For the case of NGC891 (see Fig.~12) one can see that, as 
expected, HII regions dominate the dust emission at shorter wavelengths. 
The really interesting feature of this diagram is the predominance of the UV 
heating in the submm regime. As explained by Tuffs \& Popescu \cite{tufpop03},
this can be understood as follows:  the
coldest grains are those which are in weaker radiation fields, either
in the outer optically thin regions of the disk, or because they are
shielded from radiation by optical depth effects. In the first
situation the absorption probabilities of photons are controlled by
the optical properties of the grains, so the UV photons will dominate
the heating. The second situation arises for dust associated with the
young stellar population, where the UV emissivity far exceeds the
optical emissivity.

\begin{figure}[htb] 
\includegraphics[scale=0.75]{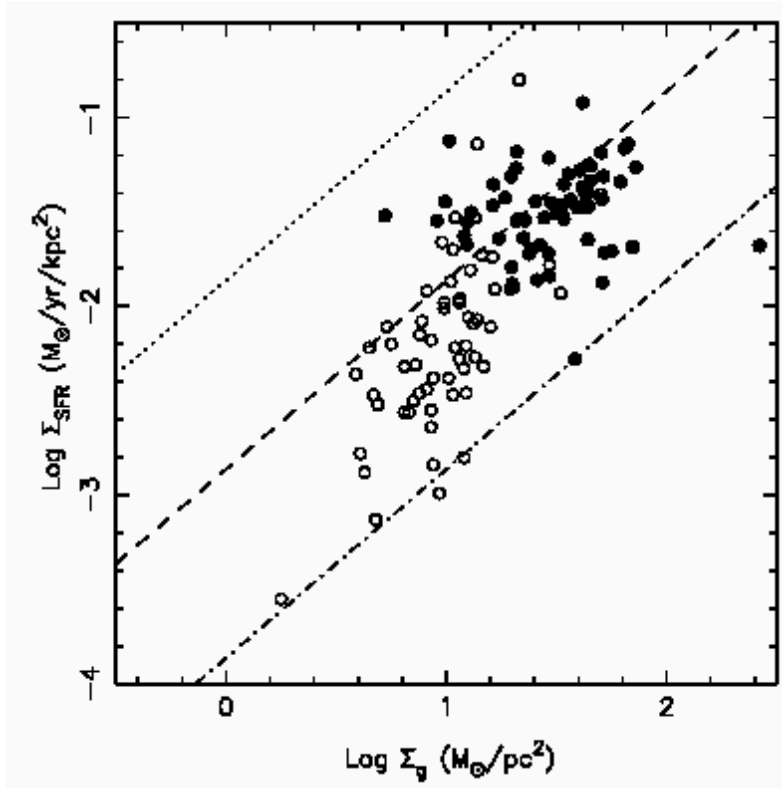} 
\caption{Star-formation rate surface density predicted from
  the SED modelling (solid circles) of a submm selected sample and from 
the H$\alpha$ measurements (open circles) of a sample of normal galaxies,  
as a function of gas surface density (Misiriotis et al. \cite{mis04}). 
The dotted, dashed and dot-dashed lines
correspond to star formation efficiencies of 100$\%$, 10$\%$ and 1$\%$ in 
$10^8$ yr.}
\end{figure}

\subsection{4.4 Application to statistical samples}

Perhaps the biggest challenge faced by these models is to make them tractable
for application to statistical samples of galaxies, while at the same time
retaining their fundamental predictive power. A first step in this direction 
was made by Misiriotis et al. \cite{mis01} who applied the model of Popescu et
al. \cite{pop00} to further edge-on galaxies modelled in the optical by 
Xilouris et al. \cite{xil99}. More recently, Misiriotis et al. \cite{mis04}
used the same model, in simplified form, to fit the FIR SEDs of a submm 
selected sample of local universe spiral galaxies. As an example of what one 
can derive from such a study we show the derived star-formation rate surface 
density as a function of gas surface density (see Fig.~13).

\section{5. A closer look at the role of clumps in shaping the SEDs of galaxies} 

We have seen that clumps of dust with embedded young stars are needed to
account for the FIR/submm SEDs of normal galaxies. We will refer to such clumps
as ``active clumps'' (with embedded sources). Their main effect is to provide a
warm component of emission to the FIR/submm SEDs.

\begin{figure}[htb] 
\includegraphics[scale=0.98]{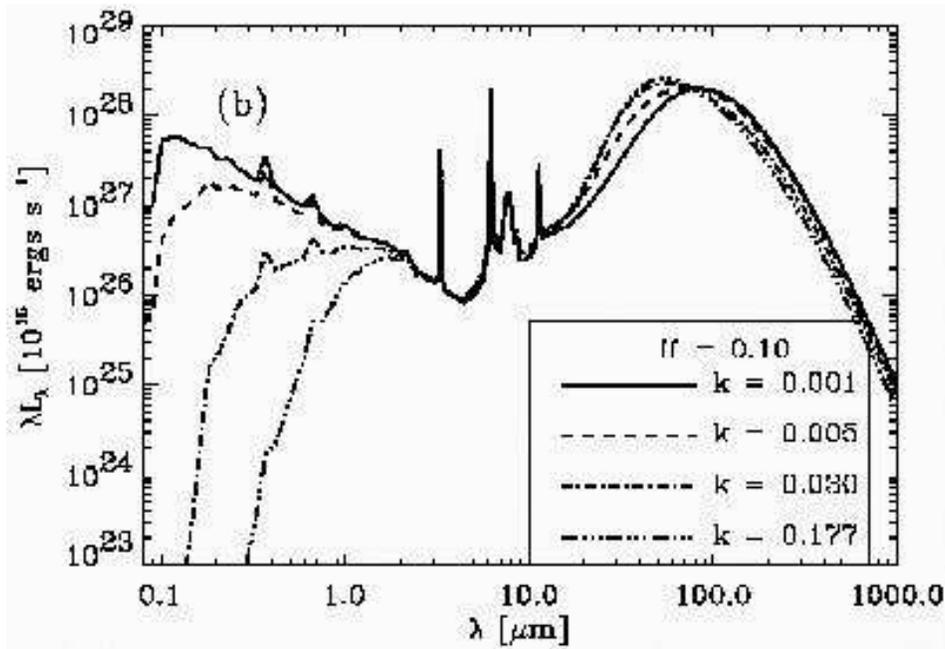}
\caption{The effect of ``passive'' clumps on the FIR/submm SEDs, 
  taken from Misselt et al. \cite{miss01}. Model SEDs are for a range of 
density ratios $k$ between the diffuse and clumpy media and for a filling 
factor ff = 0.1. All models are calculated assuming a spherical shell 
geometry. It is obvious that the effect of increasing the degree of 
clumpiness (decreasing k) is to make the SEDs cooler.}
\end{figure}

An alternative concept for clumps is to consider them as passive molecular
clouds with no embedded stars, such as those introduced by Witt \& Gordon
\cite{witgor96}, or Bianchi et al. \cite{bia00b}, and further explored by Witt
\& Gordon \cite{witgor00} and by Misiriotis \& Bianchi \cite{misbia02}.  
We will refer to such clumps
as ``passive clumps'', since they are randomly distributed with no small 
scale (of a few pc) correspondence with the young stars. The effect of passive 
clumps on the FIR/submm SED is exactly the opposite of the effect of the active
clumps, namely they provide cooler FIR SEDs. This is illustrated in Fig.~14,
which shows results from a calculation made by Misselt et al. \cite{miss01} 
for a spherical shell geometry with varying degrees of clumpiness. 
The reason why the grains are colder when the dust is distributed in passive
clumps is simply that light is no longer reaching grains in the body of the
clumps. Of course this effect only occurs when the clumps are optically
thick. A medium of optically thin clumps would make no measurable difference 
to the FIR SED compared to the same dust homogeneously distributed.

The effect of the clumps on the UV/optical/NIR SEDs is also dependent on
whether the clumps are ``active'' or ``passive''.
The effect of ``active'' clumps on the attenuation characteristics of spiral 
galaxies has been studied by Bianchi et al. \cite{bia00b}, Misiriotis \&
Bianchi \cite{misbia02} and Tuffs et al. \cite{tuf04}. Their effect has been 
also considered by Charlot \& Fall
\cite{chafal00} and, in the context of starburst galaxies, by Dopita et
al. \cite{dop05}. 
Unlike the effect of diffuse dust in a disk galaxy, the attenuation produced 
by ``active'' clumps does not strongly depend on the inclination of the 
galaxy. Furthermore, the wavelength dependence of the attenuation is no longer
primarily determined by the optical properties of the grains, but instead by 
the different blocking action of the clumps on the photons emitted by stars of
different ages, and thus masses. This effect arises because longer lived stars
are typically to be found at larger distances from their parent molecular
clouds. Furthermore, the introduction of dust in the form of active clumps will
increase the global attenuation of a galaxy more than if the same dust had been
added to the diffuse interstellar medium. This can be physically attributed to 
the placing of the clumps in the direct vicinity of the UV emitting stars.

The effect of ``passive'' clumps on the attenuation characteristics of spiral 
galaxies has been studied by Bianchi et al. \cite{bia00b}, 
Kuchinski et al. \cite{kuc98}, Misiriotis \& Bianchi \cite{misbia02} and 
Pierini et al. \cite{pie04}. 
Their main effect is to lower the global attenuation of the galaxy 
and only to a lesser degree to change the dependence  of attenuation on
wavelength. Details of the dependence of the attenuation on
inclination and wavelength appear to depend strongly on the parameterisation 
of the clumps, for example whether the clumps have a fixed density or have a 
fixed density contrast to the diffuse medium (see Bianchi et al. \cite{bia04}).

In summary, the major effects of clumps on SEDs are as follows:
\begin{itemize}
\item ``active'' clumps (with embedded sources)
\begin{itemize}
\item  tend to make the FIR/submm colours warmer
\item tend to increase the attenuation of the UV/optical emission 
\end{itemize}
\item ``passive'' clumps
\begin{itemize}
\item  tend to make the FIR/submm colours cooler
\item  tend to decrease the attenuation of the UV/optical emission
\end{itemize}
\end{itemize}

In practice, the description of real galaxies requires the presence of both 
``passive'' and ``active'' clumps. Indeed, the same physical interstellar 
cloud can act either as a passive or an active clump, depending on the 
observed wavelength. For example, molecular clouds will behave as ``active'' 
clumps in the UV, due to their strong spatial correlation with young stars,
and as ``passive'' clumps in the optical/NIR, due to the smooth distribution 
of the old stellar population. This has an interesting consequence on the 
slope of the attenuation curve of spiral galaxies, namely that they will be 
steeper than predicted by homogeneous models. This may be important to bear 
in mind when considering the contribution of spiral galaxies to the Madau's 
plot. In the dust emission, clumps with embedded sources provide the short 
wavelength luminosity. Passive clumps are unimportant in terms of luminosity, 
but will increase the amplitude of the submm emission deep in the 
Rayleigh-Jeans regime.

Of course the results we have described are for a particularly simple 
situation, that of a two phase medium (diffuse and clumps). In reality the 
interstellar medium is turbulent, presenting a whole distribution of column 
densities along different lines of sight. The effect of this more realistic
structure on attenuation has been considered by Fischera et al. \cite{fis03}, 
 Fischera \& Dopita \cite{fisdop04}, \cite{fisdop05}.

\section{6. Modelling the attenuation of stellar light in galaxies}

Traditionally, the effect of dust in attenuating stellar light
in galaxies has been quantified purely empirically
by statistical analysis of the variation of optical surface brightness with 
inclination. However, this variation is small in comparison with the
scatter, so that this method is not useful for estimating the attenuation
of light in individual galaxies. For this reason, there has always been
a strong incentive to develop models for the attenuation of stellar light
in galaxies. With the development of radiative transfer calculations
on high speed computers, such models have become a practical possibility.
The problem, however, has been to identify realistic geometries for stars and 
dust, since, as demonstrated in this review, these are not well constrained 
by optical data alone. Consequently, until very recently, models for
attenuation have largely concentrated on evaluating the effect different 
choices for the geometries of stars and dust would have on the
attenuation. The geometry was varied to investigate the effect of the
inclusion or non-inclusion of different structures, such as clumps or bulges,
as well as simply to investigate the effects of varying the parameters of
structures, such as scalelengths and scaleheights of disks of dust and stars.
 
This type of work was done by Bianchi \cite{bia96},
Ferrara et al. \cite{fer99}, Baes \& Dejonghe \cite{baedej01} and 
Pierini et al. \cite{pie04}.  For example 
Ferrara et al. provided
an atlas of dust attenuation calculated as a function of geometry, 
inclination, dust optical depth, and extinction properties. This atlas is 
available in electronic format. As an application they compared the predicted 
variation of apparent magnitude with inclination with the observed one. 
Pierini et al. \cite{pie04} investigated the effect
of the dust being distributed in ``passive'' clumps (as defined in Sect.~5),
rather than homogeneously. In this work the large scale distributions of
stars and dust and the dust-type were
 fixed, and the attenuation 
characteristics of the bulge and disk as seen through the whole 
distribution of dust were calculated separately.

The recent development of models which reproduce the observed 
optical/FIR/submm SEDs and the surface brightness distributions in both 
optical and FIR range means
that the geometry of stars and dust is no longer a free choice, but 
should conform to the solution adopted by these models. Such an approach was
taken by Tuffs et al. \cite{tuf04}, who used the SED model of Popescu et al. 
\cite{pop00} to calculate the attenuation of stellar light in spiral galaxies.
Unlike previous studies of attenuation, this model incorporates a second disk 
of dust and ``active'' clumps. These new elements affect both the inclination
and wavelength dependence of attenuation. In order to make the model applicable
to a wide range of bulge-to-disk ratios and degrees of clumpiness, separate 
calculations were done for the main stellar components: older stellar disk, 
young stellar disk and bulge, each seen through two dust disks, while the
clumpy component was treated analytically. In Fig.~15 we show examples
of attenuation curves taken from Tuffs et al. \cite{tuf04}. For all 
geometrical components there is an obvious increase in attenuation with 
increasing inclination, with the exception of the attenuation of the bulge 
when the inclination approaches the edge-on view. Also, the increase in 
attenuation with increasing inclination is stronger for larger face-on 
opacities than for lower ones, irrespective of geometry or wavelength. These 
solutions of attenuation as a function of inclination, opacity and wavelength 
are available in electronic format for general use by the community.  
These data can be used to construct attenuation curves for any combination of 
bulge-to-disk ratio and clumpiness factor. 
The effect of varying the bulge-to-disk ratio is one of the most
important factors in shaping the attenuation curves in the optical/NIR range. 
 In general, ignoring the presence of bulges can lead to a 
systematic overestimate of the opacity of disks. Also the effect of varying the
clumpiness factor is one of the most important factors in shaping the 
attenuation curves in the UV range.

\begin{figure}[htb] 
\includegraphics[scale=0.75]{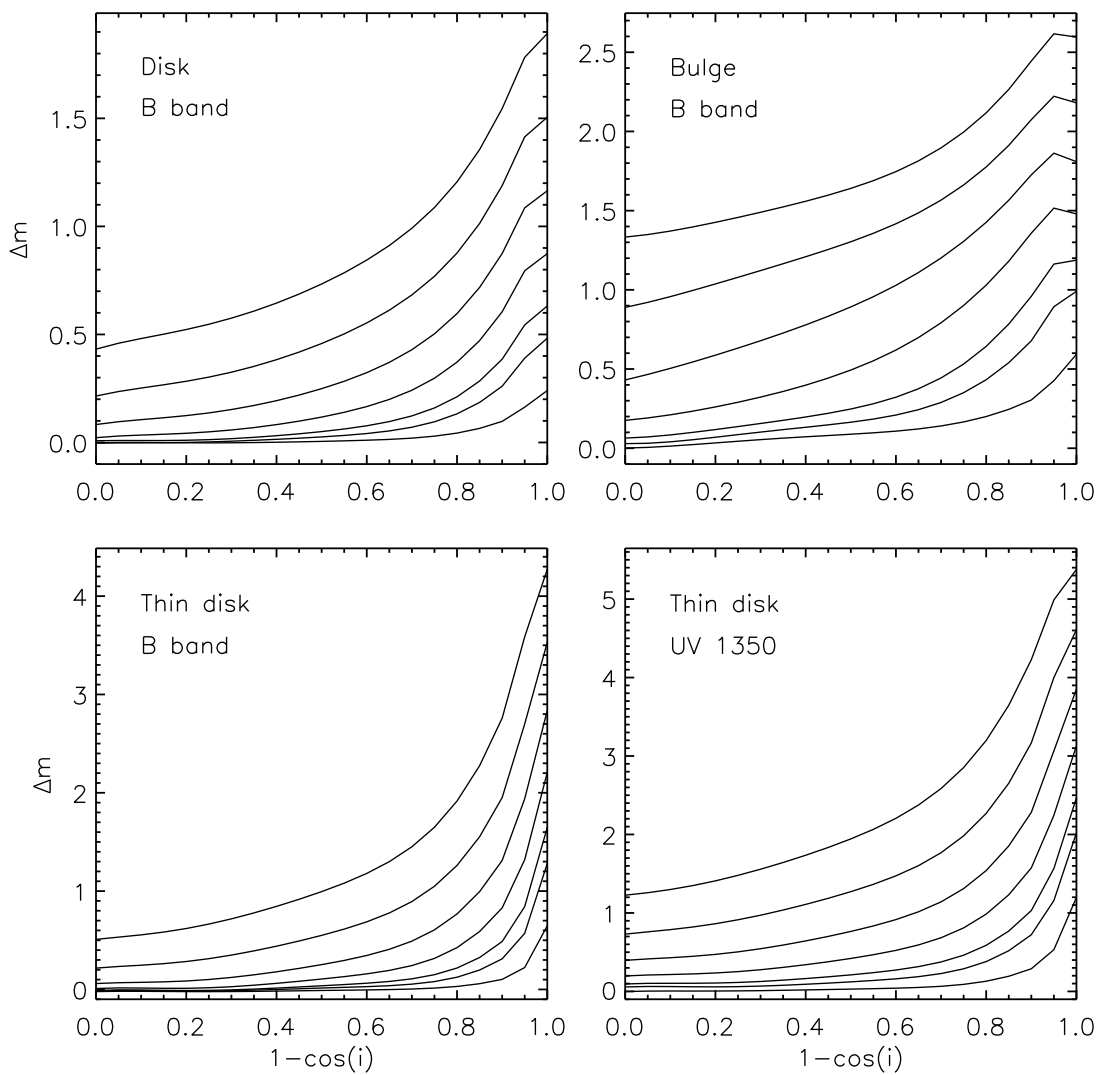} 
\caption{Examples taken from Tuffs et al. \cite{tuf04} of the dependence of 
attenuation ($\Delta m$) on inclination ($i$) for the main
stellar components their model: disk (top left), bulge(top right), and thin
disk (bottom left and right). The examples are plotted for the B band for the
disk and bulge, and both for the B band and UV 1350$\AA$ for the thin disk.
Each panel shows (from top to bottom) 7 attenuation curves, 
corresponding to central face-on B band optical depths of 
$\tau^{\rm f}_{\rm B}$: 8, 4, 2, 1, 0.5, 0.3, and 0.1. The 
face-on orientation corresponds to $1-\cos(i) = 0.0$ and the edge-on 
orientation corresponds to  $1-\cos(i) = 1.0$.}
\end{figure}
 
\section{Outlook}
Although much has already been achieved, there are clear objectives that
need to be met to make further progress in this field.
First of all, there is
still the need to make detailed modelling of nearby face-on galaxies, both in
terms of their integrated SEDs and surface-brightness distributions. 
Unlike the edge-on geometry, the face-on view also allows the young stellar 
population and associated dust to be constrained by the morphology of the 
extincted disk in the UV. Secondly, as already mentioned, we must make the 
models tractable for application to statistical samples of galaxies, while at 
the same time retaining their fundamental predictive power. For example, from
the perspective of the infrared community, we must provide a viable 
alternative to Planck curves. 

Finally, one would like to impose a physically self-consistent 
framework on these models,  in which the geometry of the stellar populations 
and of the gas and dust is calculated self-consistently in terms of the
gravitational potential of the galaxy.

\begin{theacknowledgments}
We would like to thank N.D. Kylafis, M.A. Dopita, A. Misiriotis, M. Xilouris,
J. Fischera, B.F. Madore \& H.J. V\"olk for the many discussions and fruitful 
collaboration we have enjoyed over the last years, without which this 
paper would not have been possible. N.D. Kylafis is also acknowledged for
a critical reading of this manuscript.
\end{theacknowledgments}

\end{document}